\documentclass[fleqn,10pt]{wlscirep}
\usepackage[utf8]{inputenc}

\usepackage{graphicx}
\usepackage{amsfonts}
\usepackage{amssymb}
\usepackage{amsmath}
\usepackage{newclude}
\usepackage{appendix}
\usepackage[square,comma,compress,numbers]{natbib}
\usepackage{hyperref}
\usepackage{float}
\usepackage{subcaption}
\usepackage[T1]{fontenc}

\title {Routes Obey Hierarchy in Complex Networks}

\author[1]{Attila Csoma}
\author[1]{Attila K\H{o}r\"osi}
\author[1]{G\'abor R\'etv\'ari}
\author[1]{Zal\'an Heszberger}
\author[1]{J\'ozsef B\'ir\'o}
\author[2]{Mariann Sl\'iz}
\author[3]{Andrea Avena-Koenigsberger}
\author[4,5]{Alessandra Griffa}
\author[4,5]{Patric Hagmann}
\author[1,6,*]{Andr\'as Guly\'as}
\affil[1]{Budapest University of Technology and Economics, MTA-BME Information Systems Research Group, H-1117 Budapest, Magyar
  tud\'osok krt. 2, Hungary}
\affil[2]{E\"otv\"os Lor\'and University, Institute of Hungarian Linguistics and Finno-Ugric Studies, H-1088 Budapest, M\'uzeum
  krt. 4/A, Hungary}
\affil[3]{Indiana University, Psychological and Brain Sciences, Bloomington IN, 47405, USA
}
\affil[4]{Department of Radiology, Centre Hospitalier Universitaire Vaudois (CHUV) and University of Lausanne (UNIL), Lausanne, Switzerland}  
\affil[4]{Signal Processing Laboratory 5 (LTS5), \'Ecole Polytechnique F\'ed\'erale de Lausanne (EPFL), Lausanne, Switzerland}  
\affil[6]{A. Guly\'as was supported by the J\'anos Bolyai Fellowship of the Hungarian Academy of Sciences}
\affil[*]{Corresponding author: Andr\'as Guly\'as (gulyas@tmit.bme.hu)}

\begin{abstract}

  Various hypotheses exist about the paths used for communication
  between the nodes of complex networks. Most studies simply suppose
  that communication goes via shortest paths, while others have more
  explicit assumptions about how routing (alternatively navigation or
  search) works or should work in real networks \cite{noh2004random,
    csimcsek2008navigating, adamic2001search, ling2010global,
    watts2002identity, boguna2009navigability,
    kleinberg2000navigation}. However, these assumptions are rarely
  checked against real data
  \cite{milgram1967small,dodds2003experimental}. Here we directly
  analyze the structure of operational paths using real measurements.
  For this purpose we use existing and newly created datasets having
  both the topology of the network and a sufficient number of
  empirically-determined paths over it. Such datasets are processed
  for air transportation networks, the human brain, the Internet and
  the fit-fat-cat word ladder game. Our results suggest that from the
  great number of possible paths, nature seems to pick according to
  some simple rules, which we will refer to as \emph{routing
    policies}. First we confirm, that the preference of short paths is
  an inevitable policy element, however the observed stretch of the
  paths suggests that there are other policies at work
  simultaneously. We identify two additional policies common in our
  networks: the ``conform hierarchy'', meaning that the paths should
  obey the structural hierarchy of the network, and the ``prefer
  downstream'' policy which promotes avoiding the network core if
  possible. Building upon these simple policies, we propose a
  synthetic routing policy which can recover the basic statistical
  properties of the operational paths in networks. Our results can be
  helpful in estimating the reaction of complex systems for stress
  coming from the outside more accurately than the shortest path
  assumption permits.

\end{abstract}

\begin{document}

\flushbottom
\maketitle

\vspace{-0.5cm}

\begin{figure}[ht]
\center
      \includegraphics[width=0.32\textwidth]{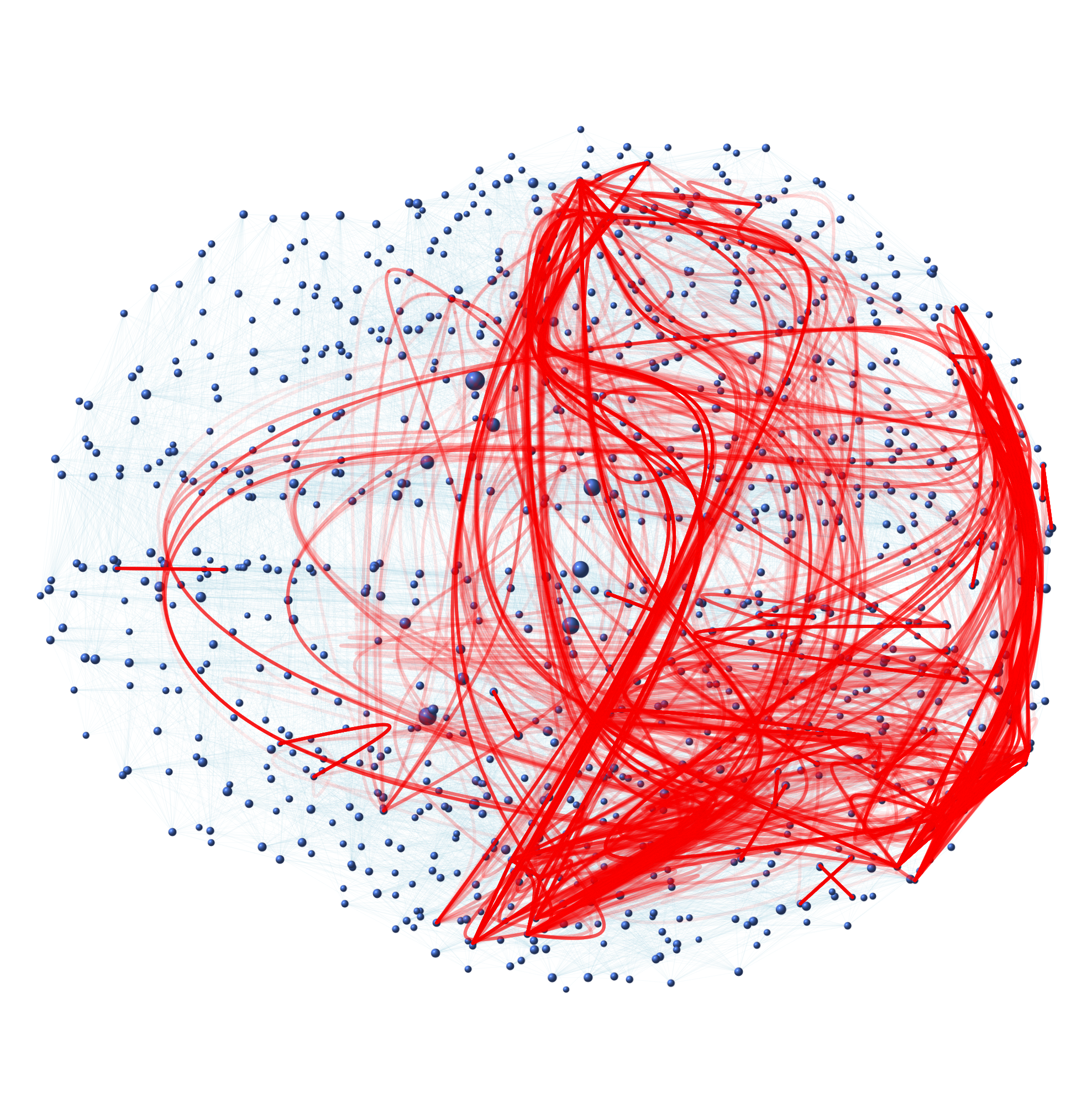}
      \put(-150.5,142.5){\large \textbf{a}}
      \includegraphics[width=0.32\textwidth]{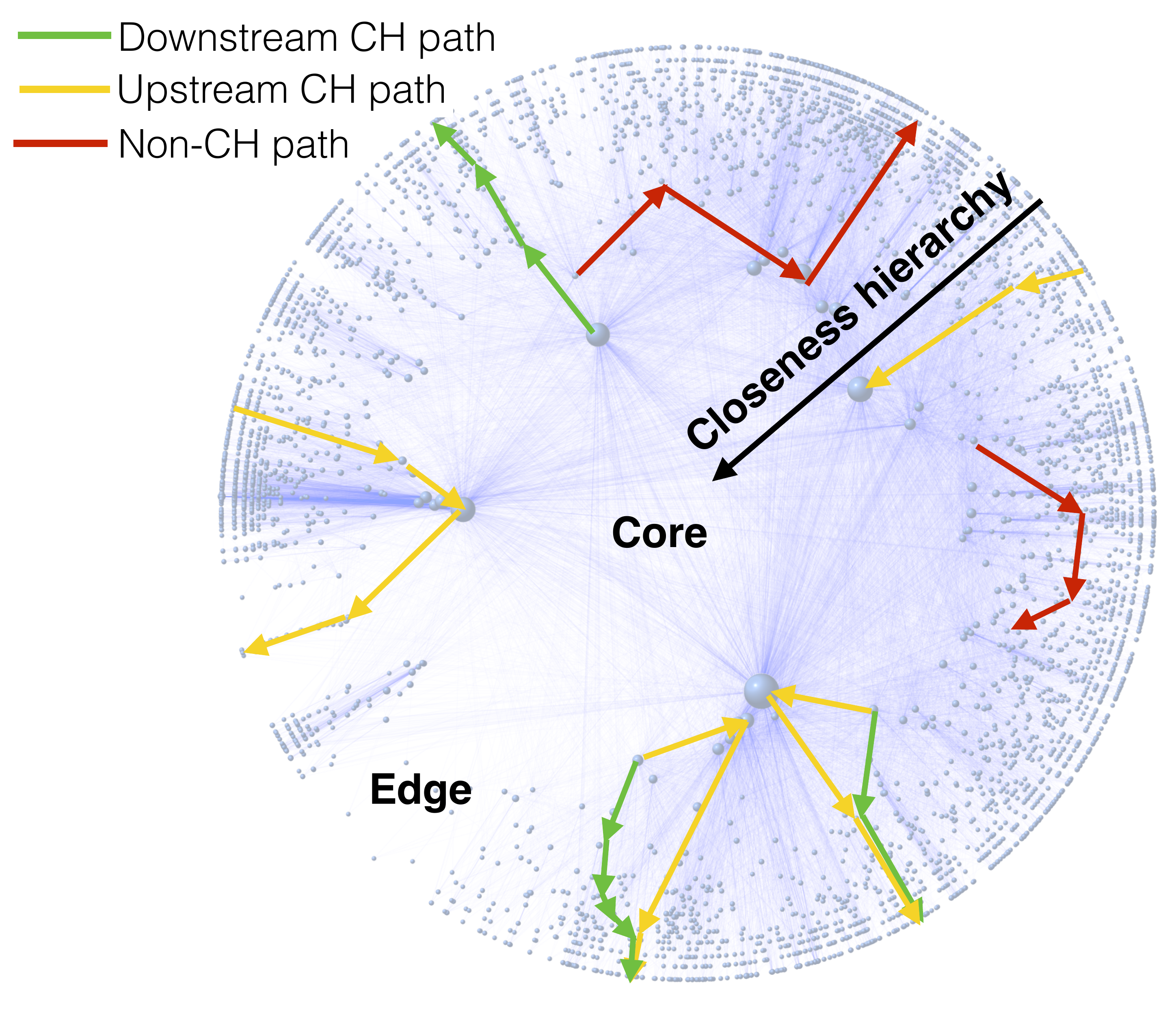}
      \put(-150.5,142.5){\large \textbf{b}}
      \caption{Visualization of the empirical paths in the human brain
        (panel \textbf{a}). The illustration of paths conforming to
        the policies identified by our measurements is shown in panel
        \textbf{b}. A path is hierarchically conform (CH) if does not
        contain a large-small-large pattern forming a `` valley''
        anywhere in its closeness centrality sequence (green and
        yellow paths in panel \textbf{b} ). An upstream path contains
        at least one step towards the core of the network (yellow
        paths), while in downstream paths the closeness centrality
        monotonically decreases (green paths).  The underlying faded
        network in panel \textbf{b} is only for illustration
        purposes. \label{fig:1}}
\end{figure}

\begin{figure}[ht]
  \centerline{
    \begin{subfigure}[b]{0.5\textwidth}
      \hspace{-0.45cm}
      \begin{tabular}{|c|c|c|c|c|}
        \hline
        Network & Airport \ & Intern. \ &  Brain & fit-fat-cat\\
        \hline
        \# Nodes & 3433 & 52194 & 1015 & 1015 \\
        \hline
        \# Edges & 20347  & 117251 & 12596.2 & 8320\\
        \hline
        Avg.\ deg.\ & 11.85 & 4.49 & 24.82 & 16.39\\
        \hline
        Avg. clust.\ & 0.64 & 0.32 & 0.42 & 0.44\\
        \hline
        Avg.\ dist.\ & 3.98 & 3.93 & 2.997 & 3.52\\
        \hline
        Diam.\ & 13 & 11 & 6.4 & 9\\
        \hline
        \# Emp. paths\ & 13722 & 2422001 & 394072 & 2700\\
        \hline
        Path avg. dist.\ & 4.67 & 4.21 & 4.16 & 3.82\\
        \hline
      \end{tabular}
      \caption{\label{fig:nettable}}
    \end{subfigure}
    \begin{subfigure}[b]{0.26\textwidth}
      \center
      \includegraphics[width=\textwidth]{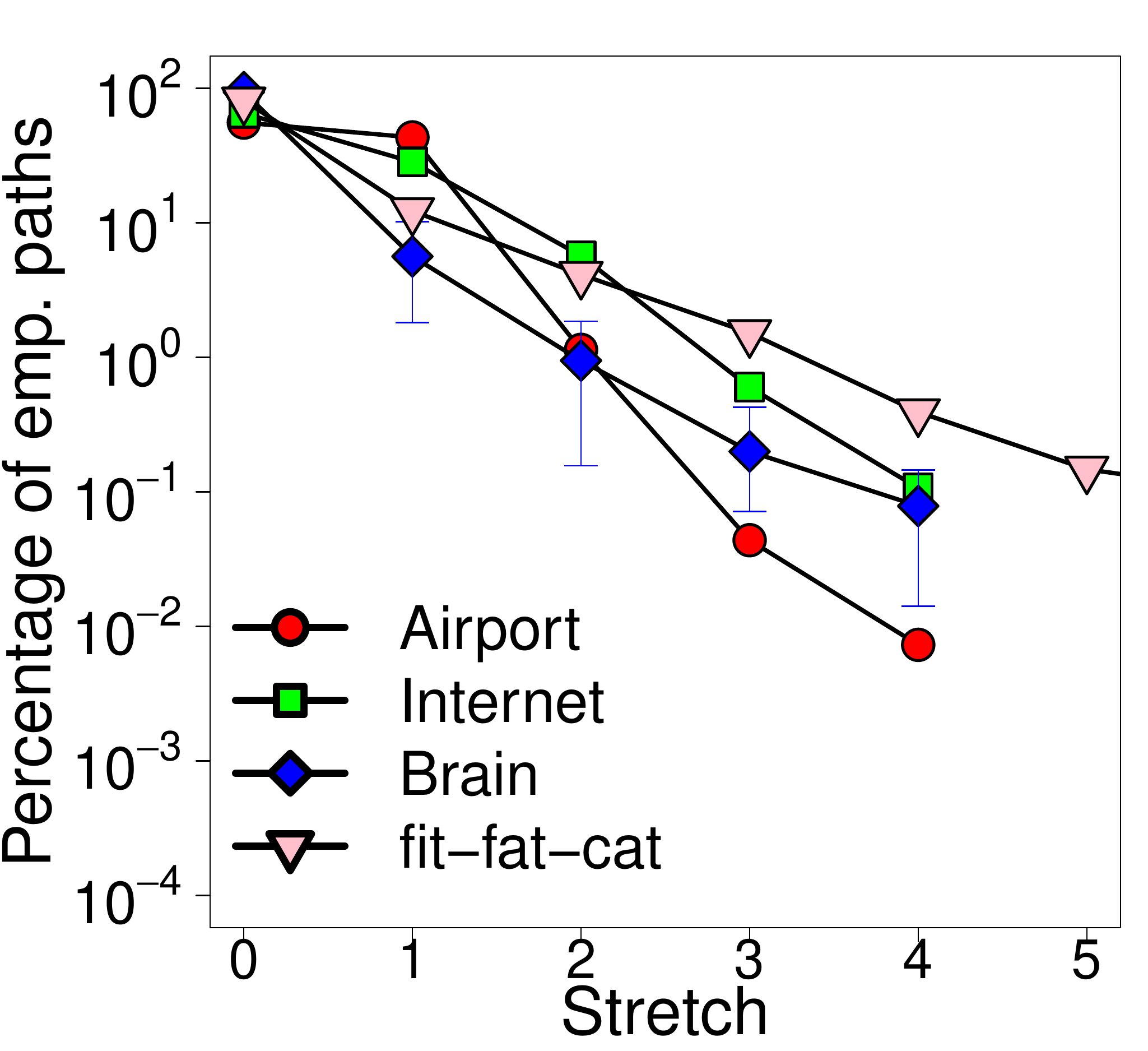}
      \caption{\label{fig:stretch}}
    \end{subfigure}
  }
  \caption{Properties of networks and stretch in empirical
    paths. Panel $(a)$ shows the basic structural properties of our
    networks and paths we have analyzed. The stretch of the empirical
    paths with respect to their shortest counterparts is shown on
    panel $(b)$. While most of the empirical paths exhibit zero
    stretch (confirming the shortest path assumption), a large
    fraction (20-40\%) of the paths is ``inflated'' even up to 4-5
    hops. The plot indicates a stunning resemblance in the
    distribution of path stretch in our networks. }
   \label{fig:2}
\end{figure}

The implicit ``shortest path'' assumption, meaning that the used
communication path in a network is the one with the shortest length,
seems to dominate the network science community and most of the
fundamental network metrics (diameter, average path length, centrality
metrics, \cite{newman2010networks} etc.) are computed using this
assumption. There are other works supposing various models
\cite{noh2004random}, network metrics (e.g. degree, centrality,
congestion, homophily \cite{csimcsek2008navigating, adamic2001search,
  ling2010global}) and hidden structures (e.g. hidden hierarchies and
metric spaces \cite{watts2002identity, boguna2009navigability,
  kleinberg2000navigation}) guiding path selection. The contribution
of these works is considerable in the modeling and understanding of
simple routing strategies that can recover near shortest paths without
requiring global knowledge of the topology. However, a lack of
confirmation with empirical data leaves an important question open:
What kind of paths are \emph{actually} chosen by nature in real world
networks?
Here we approach the question of path selection in networks from this
lacking empirical angle. Using existing and newly created datasets of
the traffic flow on real world networks (see Methods), we compare the
topology of the networks to the structure of empirically-determined
paths extracted from these datasets. From this comparison we infer common
characteristic rules of path selection in different networks, which we
call \emph{routing policies}. Our study here presents the analysis of
empirically-determined paths in air transportation networks, the
Internet, the fit-fat-cat word morph game, and empirically-inferred
paths in the human brain (see Fig.~\ref{fig:1}\textbf{a} for an
illustration). For the remainder of this text we will refer to
empirically-determined and inferred paths as empirical paths. The main
topological features of our networks and the statistics of the
empirical paths are shown in Fig.~\ref{fig:nettable}.

Our first finding is that traffic in networks does not necessarily
follow shortest paths. Fig.~\ref{fig:stretch} presents the stretch of
the paths which is computed as the length of the empirical path minus
the length of the corresponding shortest path having the same source
and destination pair. The figure shows a significant resemblance in
the distribution of path stretch across our networks.  While around
60-80\% of the empirical paths exhibit zero stretch, the remaining
paths show path stretch which can exceed up to 4-5 hops. From this
result two things follow. First, the plot confirms the shortest path
assumption of previous studies in the sense that most of the empirical
paths are shortest indeed. In this respect nature's routing policy
definitely ``prefers short paths''. However, the non-negligible
portion (20-40\%) of inflated paths suggests that there may be other
policies at use simultaneously.

\begin{figure}[H]
  \center
  \includegraphics[width=0.58\textwidth]{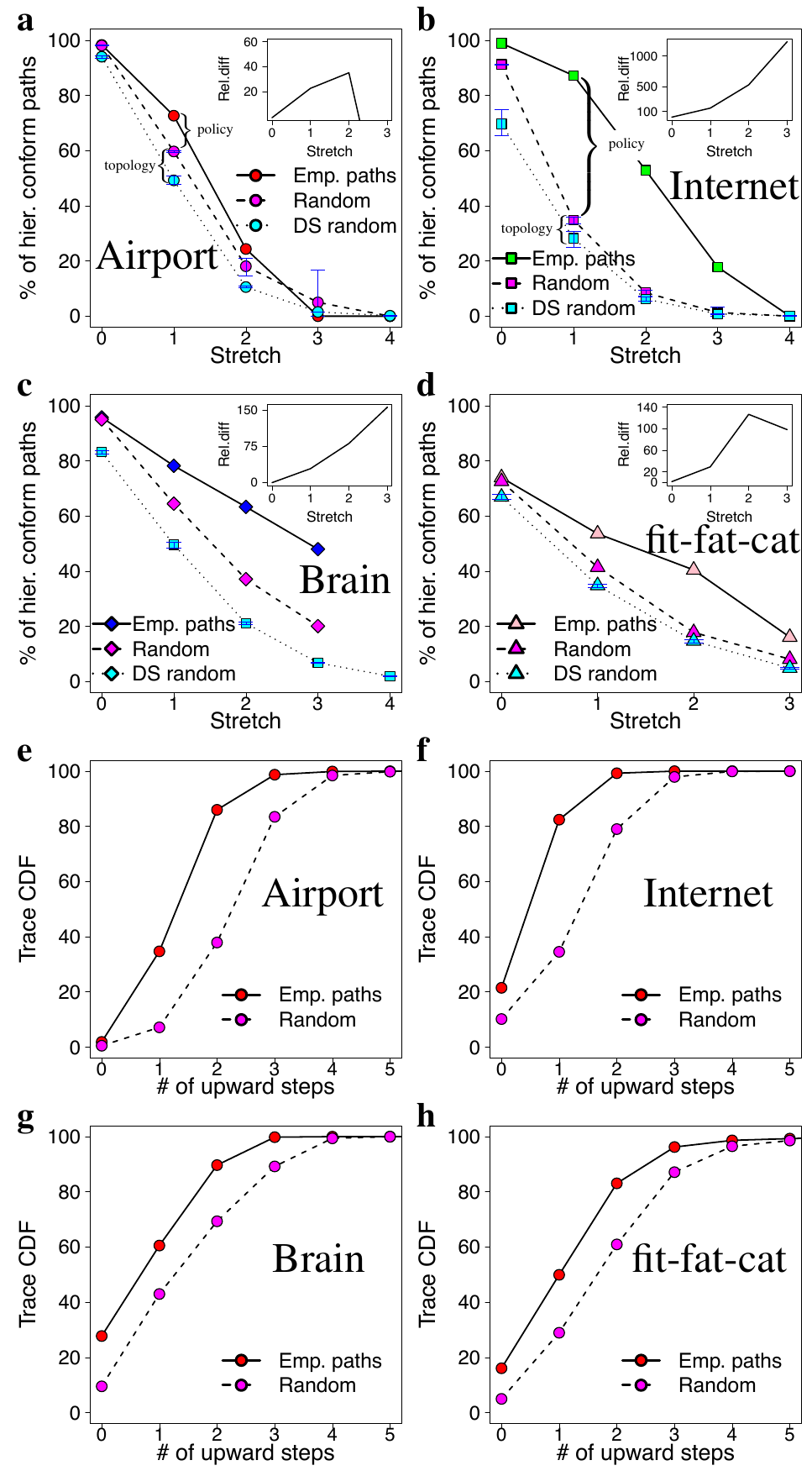}

  \caption{Identified routing policies confirmed by our measurement
    data. Panels \textbf{a}-\textbf{d} show the hierarchical
    conformity of the empirically-determined paths against
    stretch. The inset of the plots shows the relative difference
    between the number of CH paths in the empirical and the random
    paths. In the case of small networks there are 15-85\% more CH
    paths in the empirical traces but in the case of the large AS
    level Internet this goes up to 100-500\%. The cyan colored data in
    the plots show the number of CH paths in a randomized version of
    our networks generated with the degree sequence (DS) algorithm
    which produces exactly the same degrees for the nodes but the
    edges are completely randomized. The plots confirm that the
    topological peculiarities of real networks increase the number of
    CH paths between endpoints with respect to the DS networks (see
    the explanation brackets between the cyan and magenta colored dots
    of panel \textbf{a} and \textbf{b}). However, we argue that the
    effect of the CH policy is at least that important or even more
    fundamental (e.g. in case of the Internet).  Panels
    \textbf{e}-\textbf{h} show the cumulative distribution of upstream
    steps in the traces of our datasets. The empirical paths tend to
    avoid the core, which is reflected by the much lower number of
    upstream steps (in comparison with the randomly selected CH paths
    of the same length) before entering the downstream phase.}
   \label{fig:3}
\end{figure}

\vspace{-0.2cm}
One such policy our measurements support is the ``conform hierarchy''
(CH) policy, meaning that the used paths follow the topological
hierarchy of the network. For showing this we have computed the
closeness centrality of the nodes comprising the empirical paths
indicating which (inner or outer) parts of the network the information
flows through.  We found that most of the empirical paths do not
contain a large-small-large pattern forming a `` valley'' anywhere in
their closeness centrality sequence. This informally means that higher
level nodes do not prefer the exchange of information through their
subordinates even if there are short paths through them. On a CH path
the closeness centrality increases monotonically at first up to a
point (upstream), then starts to decrease (downstream) until it
reaches the destination, or it is just go upstream or downstream all
the way. Fig.~\ref{fig:1}\textbf{b} illustrates this graphically.  One
could argue that maybe short paths on real networks have this property
as a default, but Fig.~\ref{fig:3}\textbf{a}-\textbf{d} verify that
this is not the case. For comparison we picked random paths between
the source-destination pairs of our empirical paths with the same
stretch distribution and plotted the results for that case too. One
can see that, while the path length distribution is the same for the
two datasets, a much larger fraction of stretch-equivalent random
paths violate the CH policy.

There can be subtle differences between CH paths of similar
length. For example, a path can contain upstream then downstream steps
or downstream steps only. Recall that an upstream step goes towards
the core, while a downstream step goes towards the periphery of the
network. Is there a preference among these?  For answering this we
plotted the Cumulative Distribution Function (CDF) of CH paths with
respect to the number of upstream steps preceding the downstream
phase. For comparison we have also plotted the results of a random
policy which picks randomly from the possible CH paths of the given
length. The plots confirm that the empirical paths contain much less
upstream steps, which means that these paths try to avoid the
core. This finding adds ``prefer downstream'' as a third identifiable
policy component. See Fig.~\ref{fig:1}\textbf{b} for an illustration.

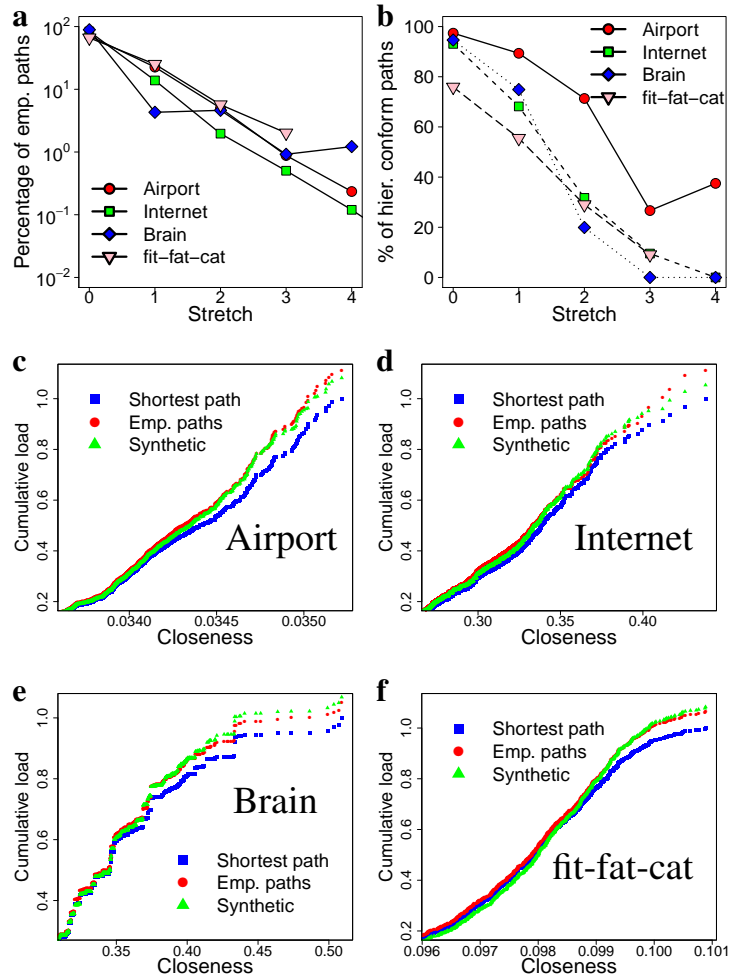
\begin{figure}[ht]
  \centerline{
    \begin{subfigure}[b]{0.27\textwidth}
      \includegraphics[width=\textwidth]{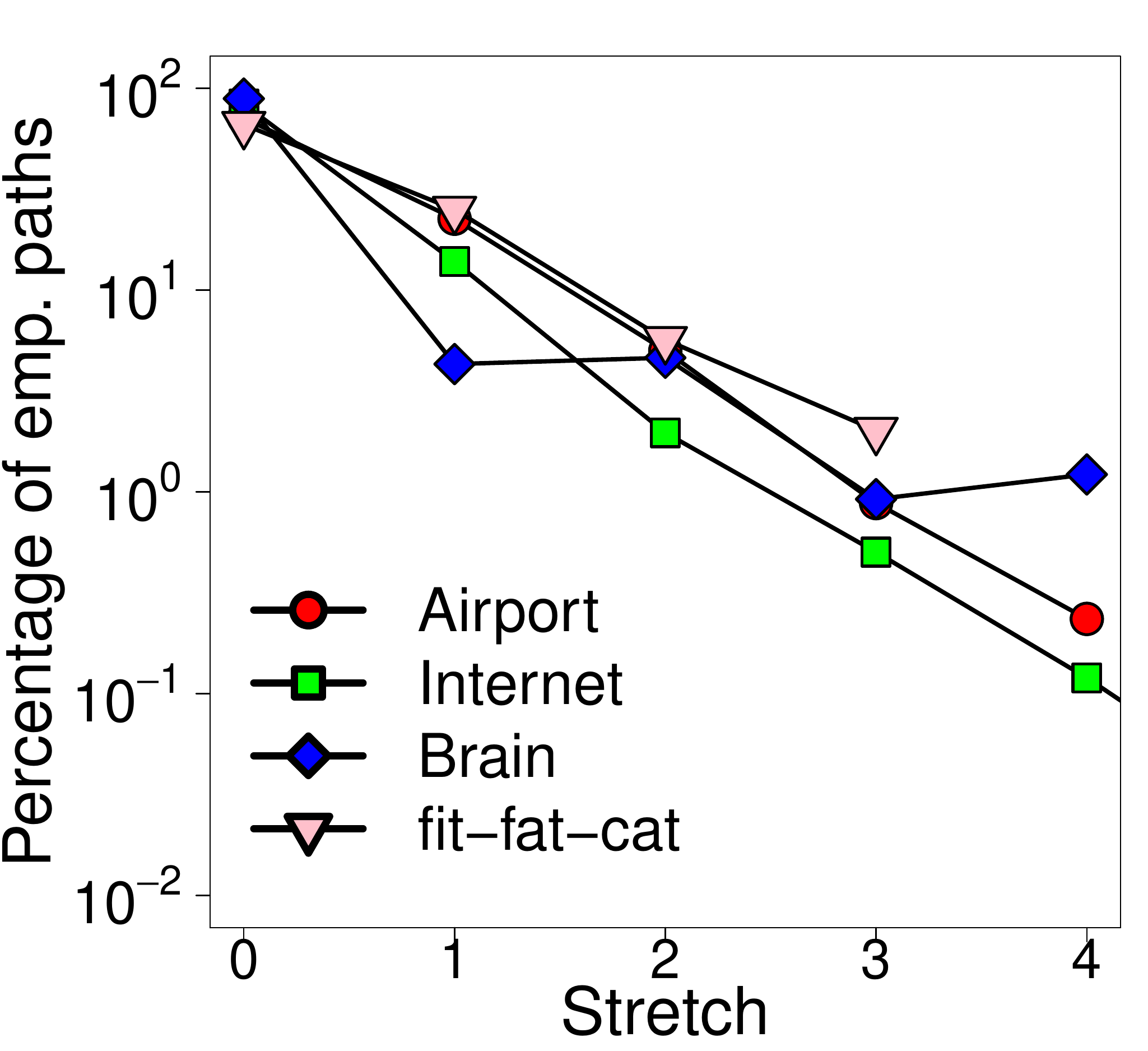}
      \put(-135.5,113.5){\large \textbf{a}}
    \end{subfigure}
    \begin{subfigure}[b]{0.27\textwidth}
      \includegraphics[width=\textwidth]{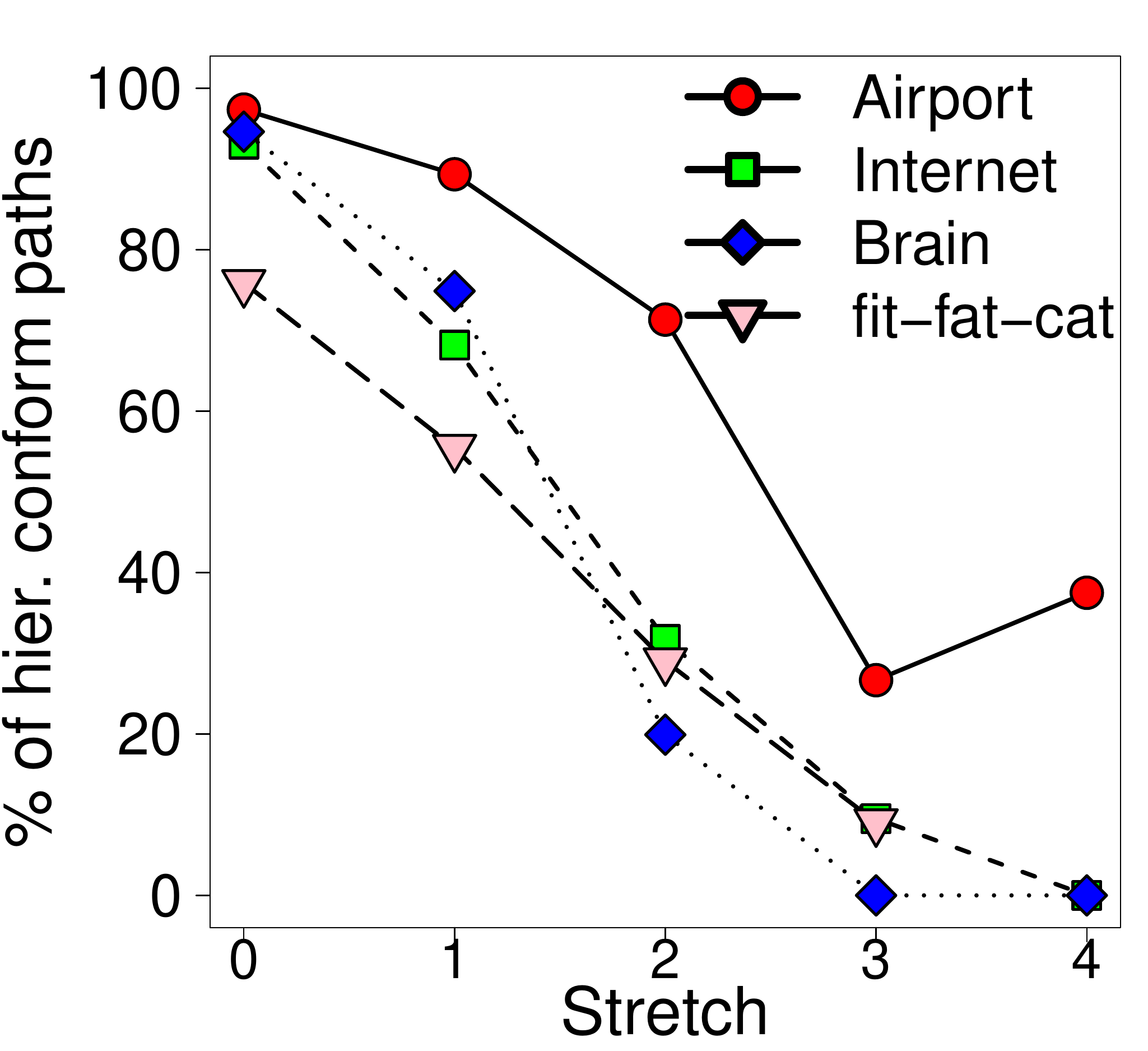}
      \put(-135.5,113.5){\large \textbf{b}}
    \end{subfigure}
  }
  \centerline{
    \begin{subfigure}[b]{0.27\textwidth}
      \includegraphics[width=\textwidth]{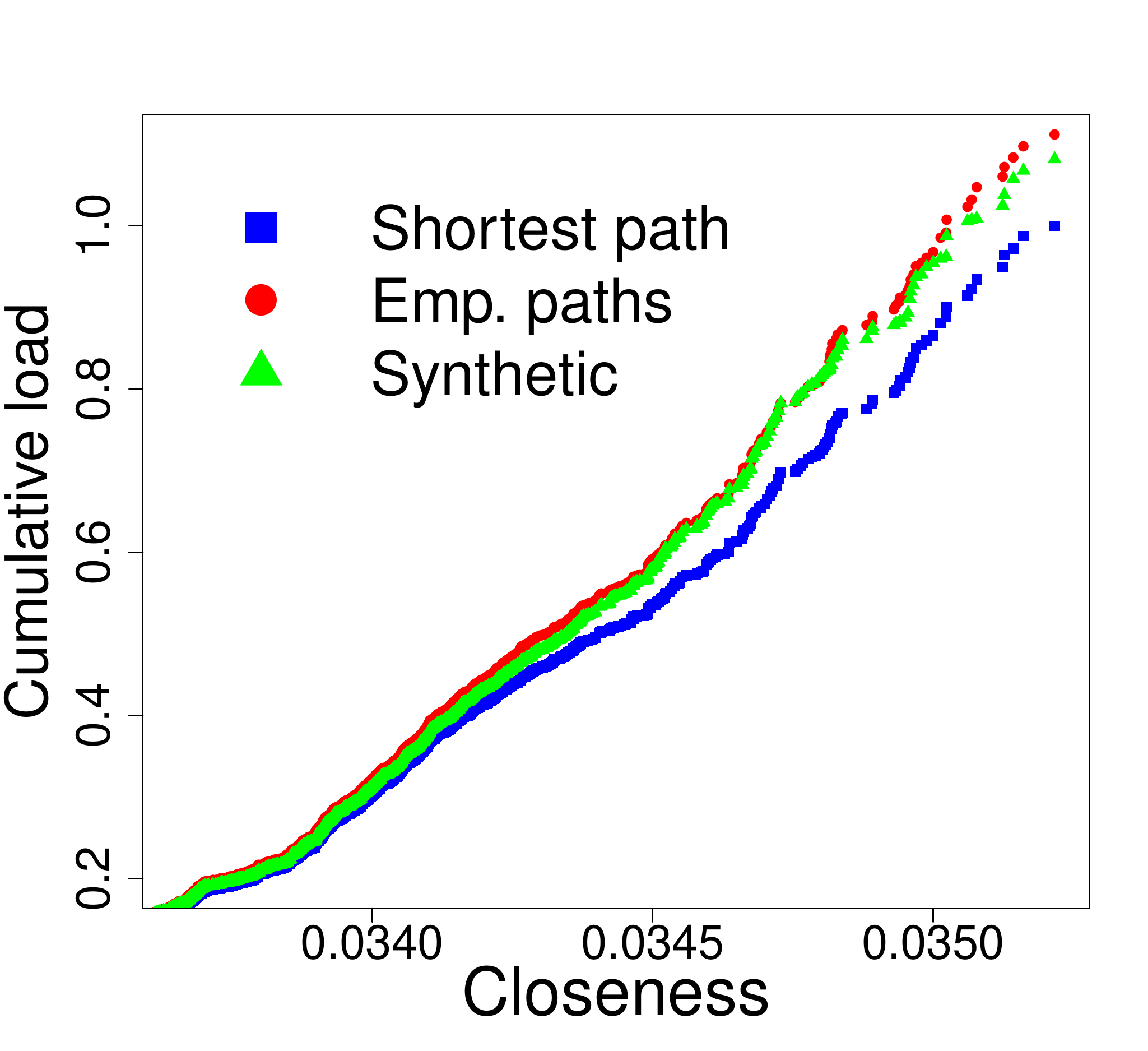}
      \put(-135.5,107.5){\large \textbf{c}}
      \put(-55,40){\Large Airport}
    \end{subfigure}
    \begin{subfigure}[b]{0.27\textwidth}
      \includegraphics[width=\textwidth]{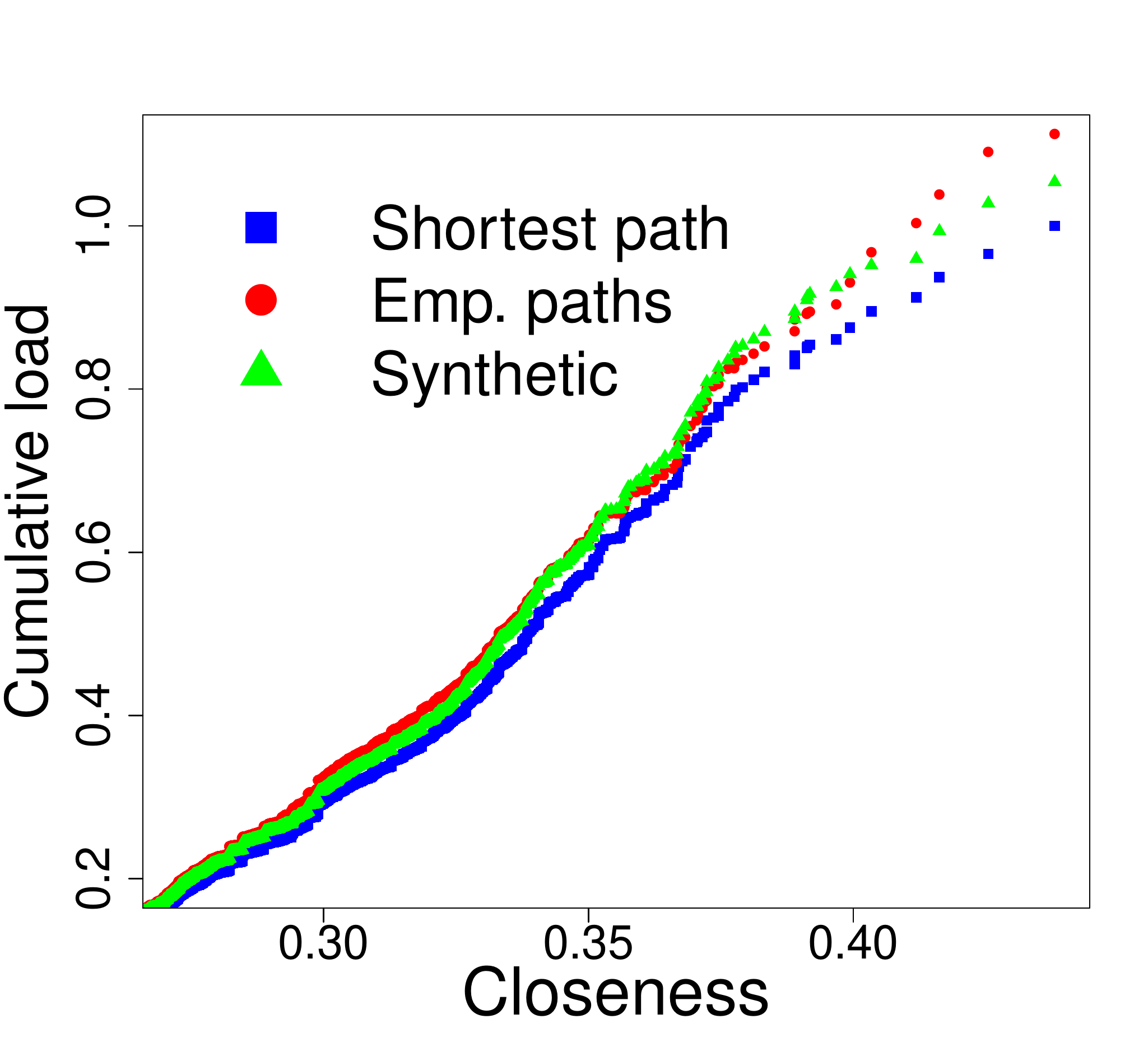}
      \put(-135.5,107.5){\large \textbf{d}}
      \put(-60.5,40){\Large Internet}
    \end{subfigure}
  }
  \centerline{
    \begin{subfigure}[b]{0.27\textwidth}
      \includegraphics[width=\textwidth]{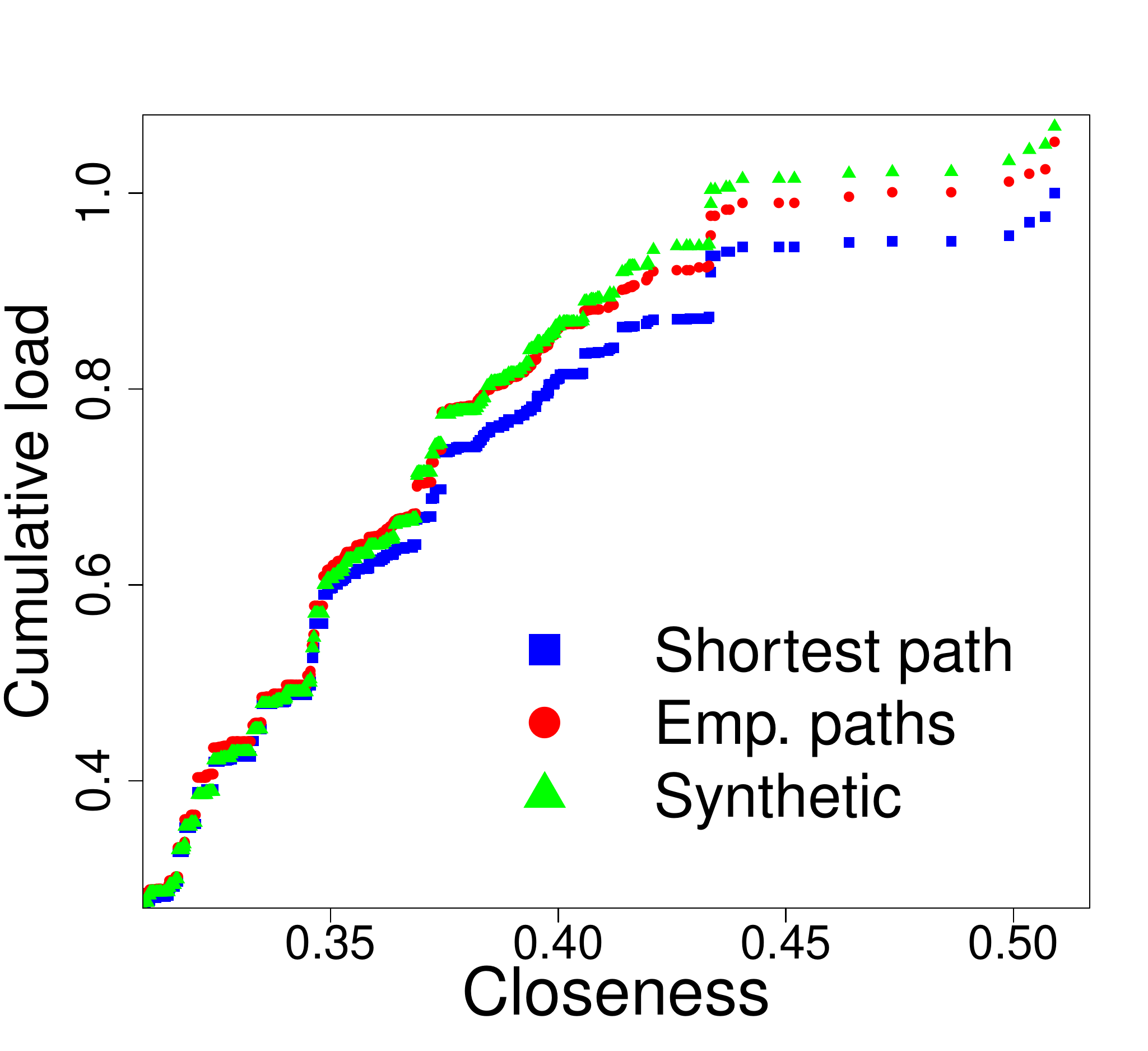}
      \put(-135.5,107.5){\large \textbf{e}}
      \put(-52,65){\Large Brain}
    \end{subfigure}
    \begin{subfigure}[b]{0.27\textwidth}
      \includegraphics[width=\textwidth]{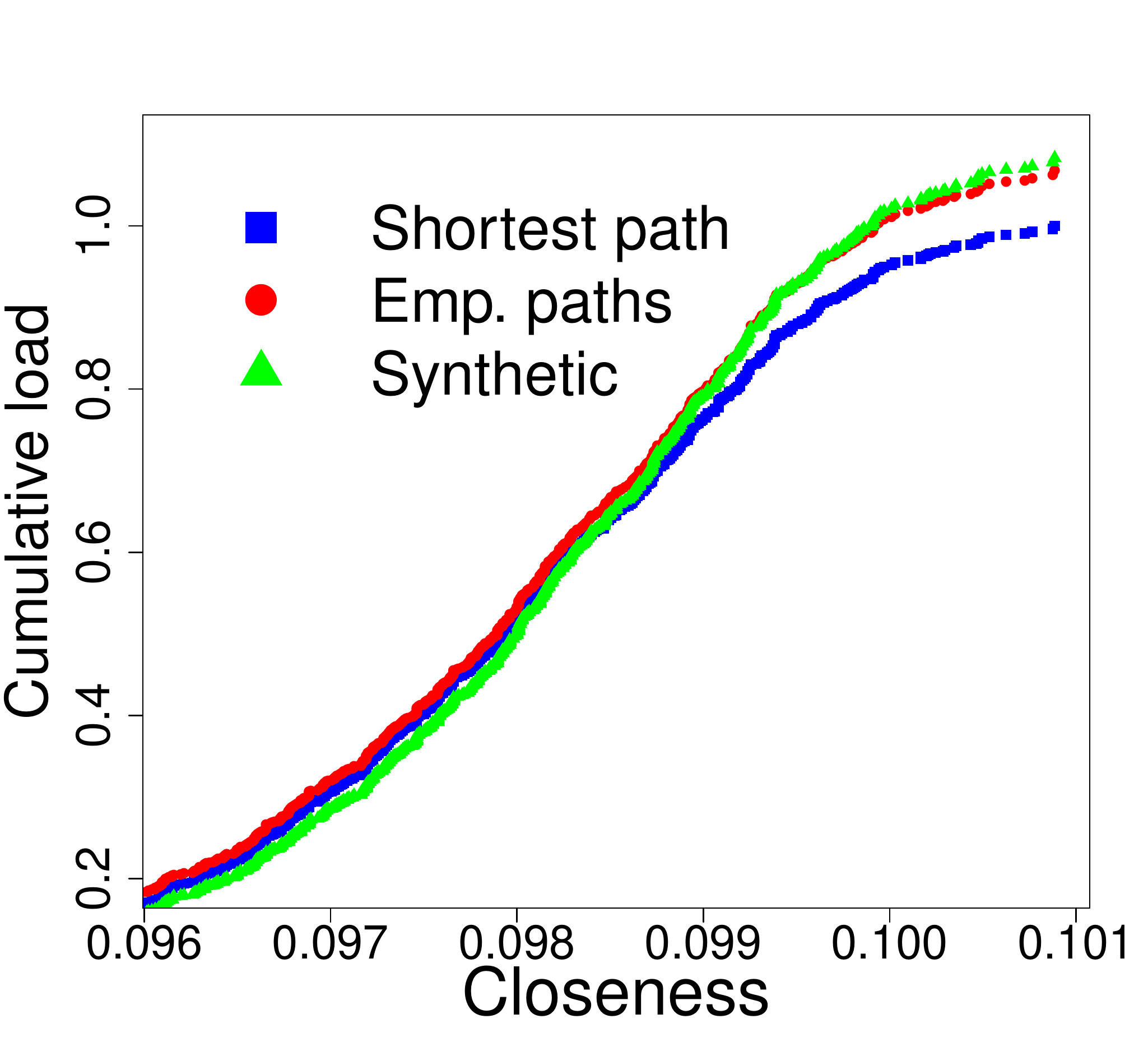}
      \put(-135.5,107.5){\large \textbf{f}}
      \put(-69,40){\Large fit-fat-cat}
    \end{subfigure}
  }
  \caption{Results of the synthetic routing policy. Our toy policy
    exhibit very realistic stretch (panel \textbf{a}) and CH
    distribution (panel \textbf{b}) (see
    Fig.~\ref{fig:2}~\textbf{a}-\textbf{b} for comparison).  Panels
    \textbf{c}-\textbf{f} present the cumulative load experienced on
    the nodes as the function of closeness, for our four datasets. The
    blue squares and the red circles show the load footprint of the
    shortest path policy and the empirical paths respectively. Due to
    the stretch of the empirical paths, the empirical plots give
    larger load that is more concentrated on the core. Our synthetic
    algorithm (green triangles) approximates this behaviour better
    than pure shortest paths. }
   \label{fig:4}
\end{figure}

Our datasets hint at the operation of the ``prefer short paths'',
``conform hierarchy'' and ``prefer downstream'' policies, with no
clear relative priorities among them. In what follows we set up a
synthetic toy routing policy which uses these components in a
prioritized fashion and, we show that using these simple ingredients
we can approximate the empirical paths much better than simple
shortest paths do. According to Fig.~\ref{fig:stretch} the prefer
shortest path policy can only have lower priority than the ``conform
hierarchy'' and the ``prefer downstream'' otherwise we would not
experience stretch at all. Since ``prefer downstream'' implies the
conform hierarchy policy, the only reasonable choice is to: prefer
hierarchically conform paths at first, then prefer downstream if there
is a downstream path and from the remaining paths prefer the short
paths. More specifically we define our synthetic routing policy to
$(i)$ use CH paths only, $(ii)$ use downstream if applicable and from
the path remaining after $(i)$ and $(ii)$ use the shortest one (if
there are still multiple choices break ties randomly). We note that
such routing policy is not unfamiliar in the literature
\cite{dodds2003information} \cite{gao2001stable}.
Fig.~\ref{fig:4}\textbf{a} shows that this simple routing policy
immediately gives very realistic path inflation, close to the stretch
computed for the real paths. What this simple algorithm cannot
reproduce is that the empirical paths sometimes violate the ``conform
hierarchy'' and the ``prefer downstream''
policies. Fig.~\ref{fig:4}\textbf{b} shows the CH distribution of the
paths generated with our synthetic routing policy over the same
network, when the closeness values are slightly randomized. This
randomization can be interpreted as simulating the case in which nodes
do not have full global information about the network and therefore,
the precise closeness values of the nodes are not known. Instead the
nodes can only have an approximate picture about the closeness
hierarchy. One can see that this modification recovers both the
stretch and CH distributions exhibited by the empirical paths (see
Fig.~\ref{fig:2}~\textbf{a}-\textbf{b}).

An immediate application of these findings is the more accurate
estimation of the network's response to outer stress (e.g. sudden load
increase).  Fig~\ref{fig:4}\textbf{c}-\textbf{f} show the load of the
nodes when carrying the traffic of the empirical paths in comparison
with shortest paths and our synthetic toy policy. One can see that
real traffic has larger volume (due to the stretch) and it is even
more concentrated in the core of the network. Our synthetic routing
policy recovers this behavior, thus, compared to a simple shortest
path policy, the synthetic policy can more accurately assess the load
footprint of real paths on the network.

\section*{Discussion}

Despite the simplicity and intuitive nature of the identified
policies, readers may have the feeling that there should be a more
simple explanation out there. For example one could imagine a
weighting of the edges over which shortest path computation will give
exactly the same paths that we find in our data.  We had two reasons
for going this way. First, our data hints for policies that are not
unfamiliar in the literature. For example, the authors of
\cite{dodds2003information} use a very similar routing algorithm for
modeling the flow of information in organizational networks, after an
in-depth analysis of such networks. Secondly, we have run experiments
to find an appropriate weighting using edge betweenness (see
Appendix \ref{app:a} for a brief listing of the results). These
experiments suggest that finding reasonable weights that are able to
generate paths matching all of our statistics is far from trivial.

\section*{Methods}

Collecting or inferring paths in networks is a non-trivial
problem. Here we list our methods for every specific networks analyzed
in our paper.

\noindent \textbf{Internet AS topology and real AS paths -- } The
Internet protocols permit the tracing of packets. We have downloaded
an AS level Internet topology and full AS level packet traces from
CAIDA (Center for Applied Internet Data Analysis, \url{www.caida.org})
recorded on 09/29/2015. The topology contains $52194$ nodes and
$117251$ connections. For this topology we had around $2.5$ million
traces.

\noindent \textbf{Air transportation network and flight travels --}
The world's flight map is available from OpenFlights
(\url{www.openflights.org}), from which the topology of the air
transportation network can be reconstructed. For a realistic
estimation of the flights used by customers, we used the Rome2Rio
(\url{www.rome2rio.com}) trip planner and generated routes between
$27444$ randomly chosen pairs of airports.  From the offered paths we
have chosen the cheapest one in the analysis in the paper. However, we
note that picking according to other parameters (lowest number of
transfers, lowest travel time) did not qualitatively change our
results. To achieve a more realistic topology we used airport
connections extracted from traceroutes to increase the accuracy of the
OpenFlight topology. The reconstructed map contained $3433$ airports
and $20347$ flights connecting them.

\noindent \textbf{fit-fat-cat word ladder game app and word chains --}
For collecting paths from word networks we have implemented a word
ladder game named ``fit-fat-cat'' for smartphones.  The goal of the
game is to transform a source word into a target word through
meaningful intermediate words by changing only one letter at a
time. The word chain fit-fat-cat is a good solution of a game with
source word fit and target word cat. These word chains, collected
anonymously from our users, can be considered as the footprints of
human navigation over the word-maze of the English language.  For the
reconstruction of the word graph we have downloaded the official
three-letter English Scrabble words from WordFind
(\url{http://www.wordfind.com/scrabble-word-list/}) and created an
edge between all the words differing only in one letter. The collected
three-letter word chains were considered as our traces. For capturing
only the ``working'' paths we have filtered out the first $20$ games
(the warming up phase) and the games taking more than 30 seconds (when
the players are not just using a known path but discover an unknown
one) of every player. After all, we have a dataset of more than 2500
paths from 100+ players.


\noindent \textbf{Human brain and estimated paths --} Getting
realistic paths from inside the human brain is extremely hard, if not
impossible. As a consequence, almost all studies in the literature
concerning path-related analysis assume shortest path signaling
paths. Taking into account the extreme non-triviality of path
estimation in the brain we ask here if we can use empirical anatomical
and functional data to infer feasible communication traces. Our
dataset comprises 40 healthy human subjects who underwent an MRI
session where Diffusion Spectrum Imaging (DSI) and resting-state
functional MRI data were acquired for each subject. DSI data was
processed following the procedures described in
\cite{hagmann2008mapping,cammoun2012mapping,daducci2012connectome},
resulting in 40 weighted, undirected structural connectivity maps
($GS$) comprising 1015 nodes, where each node represents a parcel of
cortical or subcortical gray matter, and connections represent white
matter streamlines connecting a pair of brain regions. Connection
weights determine the average density of white matter streamlines and
here only consider connections with density above 0.0001, resulting in
$GS$ with an average of $12596.2$ connections per subject. Functional
MRI data was processed following state of the art pipelines described
in \cite{murphy2009impact,power2012spurious}, yielding a BOLD signal
time-series per node, each with 276 points that were sampled every
1920 ms. The magnitude of the BOLD signal is an indicator of the
degree of neural activity at a node. Combining structural and
functional data, we infer feasible structural pathways through which
neural signals might propagate using the following process. $(i)$
Identify source-destination pairs with high statistically-dependent
brain activity. We searched for pairs of nodes such that the Pearson
correlation of the BOLD signal time series - without global regression
- was above 90\%. These nodes were used as the source-destination
pairs of our paths. $(ii)$ Determine which nodes are active at every
time-step. We say that a node is ``active'' at a given time-step if
the BOLD signal is $> \gamma$ and ``inactive'' otherwise. We construct
activity vectors for each time-step indicating which nodes were
active. Here we use $\gamma=0$ but we get qualitatively similar
results for near zero $\gamma$. $(iii)$ Construct subgraphs of active
nodes. We constructed a subgraph $GS_i$ of $GS$ for each time-step by
considering only the nodes that are active at a given time-step
$i$. $(iv)$ Define paths between source-destination node pairs. For
all of our source-destination pairs (generated in step $(i)$), we
considered the shortest path in the $GS_i$ graphs, if the path
existed. If there was multiple shortest paths between a
source-destination pair we choose one randomly. Our source-destination
traces include the paths found across all $GS_i$ subgraphs.  It is
worth noting that this method assumes that information can only
traverse active nodes. Furthermore, we are considering here a model
for large spatial and temporal scale communication in brain networks
that is not necessarily applicable to neural networks at smaller
scales. While we cannot validate with empirical data whether these
paths are actually used for the flow of neural signals, from a path
inflation perspective we can consider these paths as a lower bound on
the length of the real signaling pathways.

\bibliography{ref}

\begin{thebibliography}{10}
\expandafter\ifx\csname url\endcsname\relax
  \def\url#1{\texttt{#1}}\fi
\expandafter\ifx\csname urlprefix\endcsname\relax\def\urlprefix{URL }\fi
\expandafter\ifx\csname doiprefix\endcsname\relax\def\doiprefix{DOI }\fi
\providecommand{\bibinfo}[2]{#2}
\providecommand{\eprint}[2][]{\url{#2}}

\bibitem{noh2004random}
\bibinfo{author}{Noh, J.~D.} \& \bibinfo{author}{Rieger, H.}
\newblock \bibinfo{title}{Random walks on complex networks}.
\newblock \emph{\bibinfo{journal}{Physical review letters}}
  \textbf{\bibinfo{volume}{92}}, \bibinfo{pages}{118701}
  (\bibinfo{year}{2004}).

\bibitem{csimcsek2008navigating}
\bibinfo{author}{{\c{S}}im{\c{s}}ek, {\"O}.} \& \bibinfo{author}{Jensen, D.}
\newblock \bibinfo{title}{Navigating networks by using homophily and degree}.
\newblock \emph{\bibinfo{journal}{Proceedings of the National Academy of
  Sciences}} \textbf{\bibinfo{volume}{105}}, \bibinfo{pages}{12758--12762}
  (\bibinfo{year}{2008}).

\bibitem{adamic2001search}
\bibinfo{author}{Adamic, L.~A.}, \bibinfo{author}{Lukose, R.~M.},
  \bibinfo{author}{Puniyani, A.~R.} \& \bibinfo{author}{Huberman, B.~A.}
\newblock \bibinfo{title}{Search in power-law networks}.
\newblock \emph{\bibinfo{journal}{Physical review E}}
  \textbf{\bibinfo{volume}{64}}, \bibinfo{pages}{046135}
  (\bibinfo{year}{2001}).

\bibitem{ling2010global}
\bibinfo{author}{Ling, X.}, \bibinfo{author}{Hu, M.-B.},
  \bibinfo{author}{Jiang, R.} \& \bibinfo{author}{Wu, Q.-S.}
\newblock \bibinfo{title}{Global dynamic routing for scale-free networks}.
\newblock \emph{\bibinfo{journal}{Physical Review E}}
  \textbf{\bibinfo{volume}{81}}, \bibinfo{pages}{016113}
  (\bibinfo{year}{2010}).

\bibitem{watts2002identity}
\bibinfo{author}{Watts, D.~J.}, \bibinfo{author}{Dodds, P.~S.} \&
  \bibinfo{author}{Newman, M.~E.}
\newblock \bibinfo{title}{Identity and search in social networks}.
\newblock \emph{\bibinfo{journal}{Science}} \textbf{\bibinfo{volume}{296}},
  \bibinfo{pages}{1302--1305} (\bibinfo{year}{2002}).

\bibitem{boguna2009navigability}
\bibinfo{author}{Bogun{\'a}, M.}, \bibinfo{author}{Krioukov, D.} \&
  \bibinfo{author}{Claffy, K.~C.}
\newblock \bibinfo{title}{Navigability of complex networks}.
\newblock \emph{\bibinfo{journal}{Nature Physics}}
  \textbf{\bibinfo{volume}{5}}, \bibinfo{pages}{74--80} (\bibinfo{year}{2009}).

\bibitem{kleinberg2000navigation}
\bibinfo{author}{Kleinberg, J.~M.}
\newblock \bibinfo{title}{Navigation in a small world}.
\newblock \emph{\bibinfo{journal}{Nature}} \textbf{\bibinfo{volume}{406}},
  \bibinfo{pages}{845--845} (\bibinfo{year}{2000}).

\bibitem{milgram1967small}
\bibinfo{author}{Milgram, S.}
\newblock \bibinfo{title}{The small world problem}.
\newblock \emph{\bibinfo{journal}{Psychology today}}
  \textbf{\bibinfo{volume}{2}}, \bibinfo{pages}{60--67} (\bibinfo{year}{1967}).

\bibitem{dodds2003experimental}
\bibinfo{author}{Dodds, P.~S.}, \bibinfo{author}{Muhamad, R.} \&
  \bibinfo{author}{Watts, D.~J.}
\newblock \bibinfo{title}{An experimental study of search in global social
  networks}.
\newblock \emph{\bibinfo{journal}{science}} \textbf{\bibinfo{volume}{301}},
  \bibinfo{pages}{827--829} (\bibinfo{year}{2003}).

\bibitem{newman2010networks}
\bibinfo{author}{Newman, M.}
\newblock \emph{\bibinfo{title}{Networks: an introduction}}
  (\bibinfo{publisher}{Oxford university press}, \bibinfo{year}{2010}).

\bibitem{dodds2003information}
\bibinfo{author}{Dodds, P.~S.}, \bibinfo{author}{Watts, D.~J.} \&
  \bibinfo{author}{Sabel, C.~F.}
\newblock \bibinfo{title}{Information exchange and the robustness of
  organizational networks}.
\newblock \emph{\bibinfo{journal}{Proceedings of the National Academy of
  Sciences}} \textbf{\bibinfo{volume}{100}}, \bibinfo{pages}{12516--12521}
  (\bibinfo{year}{2003}).

\bibitem{gao2001stable}
\bibinfo{author}{Gao, L.} \& \bibinfo{author}{Rexford, J.}
\newblock \bibinfo{title}{Stable internet routing without global coordination}.
\newblock \emph{\bibinfo{journal}{IEEE/ACM Transactions on Networking (TON)}}
  \textbf{\bibinfo{volume}{9}}, \bibinfo{pages}{681--692}
  (\bibinfo{year}{2001}).

\bibitem{hagmann2008mapping}
\bibinfo{author}{Hagmann, P.} \emph{et~al.}
\newblock \bibinfo{title}{Mapping the structural core of human cerebral
  cortex}.
\newblock \emph{\bibinfo{journal}{PLoS Biol}} \textbf{\bibinfo{volume}{6}},
  \bibinfo{pages}{e159} (\bibinfo{year}{2008}).

\bibitem{cammoun2012mapping}
\bibinfo{author}{Cammoun, L.} \emph{et~al.}
\newblock \bibinfo{title}{Mapping the human connectome at multiple scales with
  diffusion spectrum mri}.
\newblock \emph{\bibinfo{journal}{Journal of neuroscience methods}}
  \textbf{\bibinfo{volume}{203}}, \bibinfo{pages}{386--397}
  (\bibinfo{year}{2012}).

\bibitem{daducci2012connectome}
\bibinfo{author}{Daducci, A.} \emph{et~al.}
\newblock \bibinfo{title}{The connectome mapper: an open-source processing
  pipeline to map connectomes with mri}.
\newblock \emph{\bibinfo{journal}{PloS one}} \textbf{\bibinfo{volume}{7}},
  \bibinfo{pages}{e48121} (\bibinfo{year}{2012}).

\bibitem{murphy2009impact}
\bibinfo{author}{Murphy, K.}, \bibinfo{author}{Birn, R.~M.},
  \bibinfo{author}{Handwerker, D.~A.}, \bibinfo{author}{Jones, T.~B.} \&
  \bibinfo{author}{Bandettini, P.~A.}
\newblock \bibinfo{title}{The impact of global signal regression on resting
  state correlations: are anti-correlated networks introduced?}
\newblock \emph{\bibinfo{journal}{Neuroimage}} \textbf{\bibinfo{volume}{44}},
  \bibinfo{pages}{893--905} (\bibinfo{year}{2009}).

\bibitem{power2012spurious}
\bibinfo{author}{Power, J.~D.}, \bibinfo{author}{Barnes, K.~A.},
  \bibinfo{author}{Snyder, A.~Z.}, \bibinfo{author}{Schlaggar, B.~L.} \&
  \bibinfo{author}{Petersen, S.~E.}
\newblock \bibinfo{title}{Spurious but systematic correlations in functional
  connectivity mri networks arise from subject motion}.
\newblock \emph{\bibinfo{journal}{Neuroimage}} \textbf{\bibinfo{volume}{59}},
  \bibinfo{pages}{2142--2154} (\bibinfo{year}{2012}).

\end{thebibliography}



\section*{Author contributions statement}

A.Cs., A.K., G.R., Z.H., J.B. and A.G. contributed to all of the
experiments and the setup of the synthetic routing.  M.S. contributed
to the fit-fat-cat experiment and A.A-K., A.Gri. P.H. contributed to the measurements
on the human brain. All authors contributed to and reviewed the
manuscript.

\section*{Additional information}

\textbf{Competing financial interests}

The authors declare no competing financial interests.


\appendix

\section{Results with betweenness based egde weights}

\label{app:a}

\begin{figure}[ht]
  \centerline{
    \begin{subfigure}[b]{0.35\textwidth}
      \includegraphics[width=\textwidth]{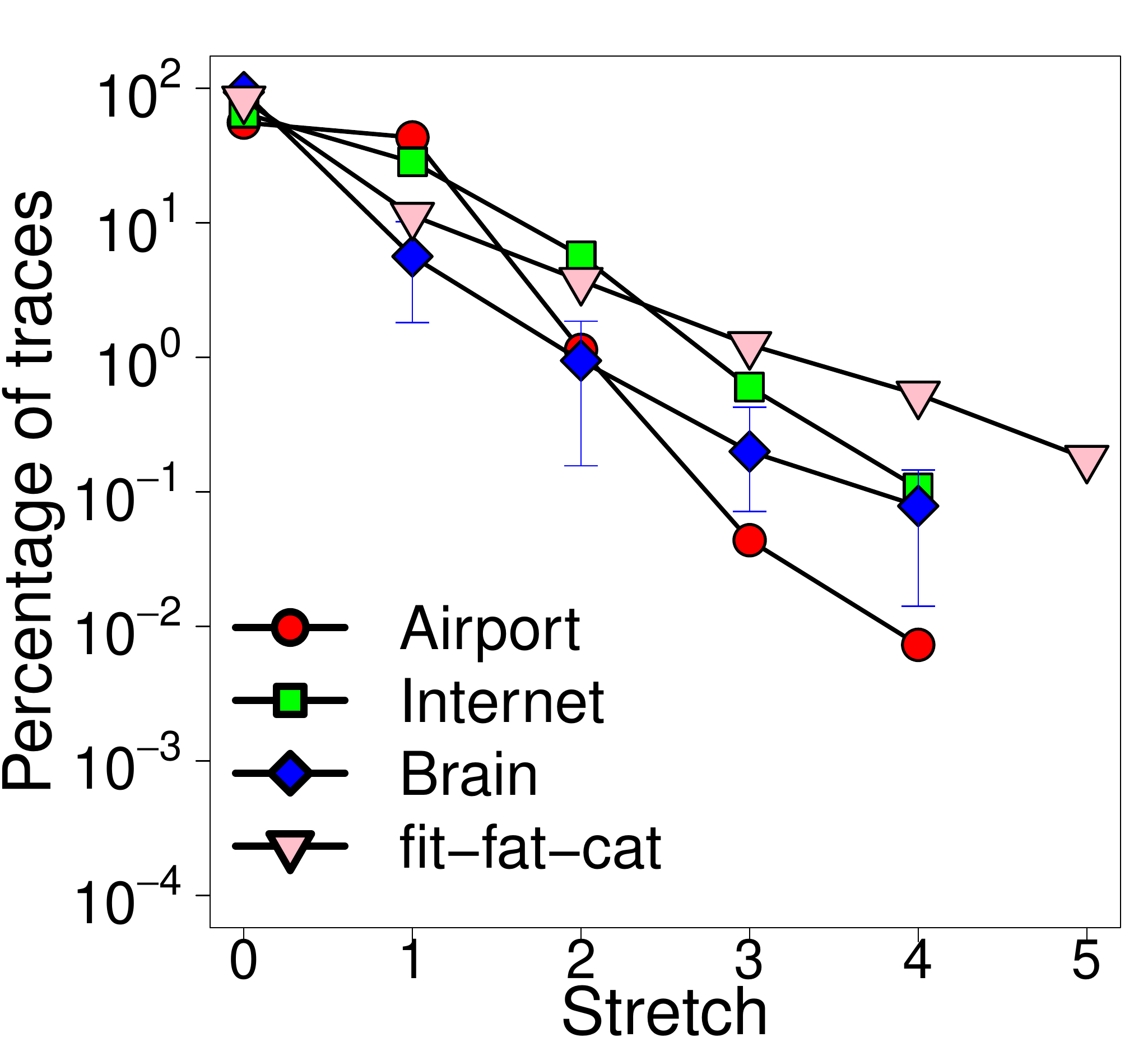}
      \put(-163.5,130.5){\large (\textbf{a})}
    \end{subfigure}
    \begin{subfigure}[b]{0.35\textwidth}
      \includegraphics[width=\textwidth]{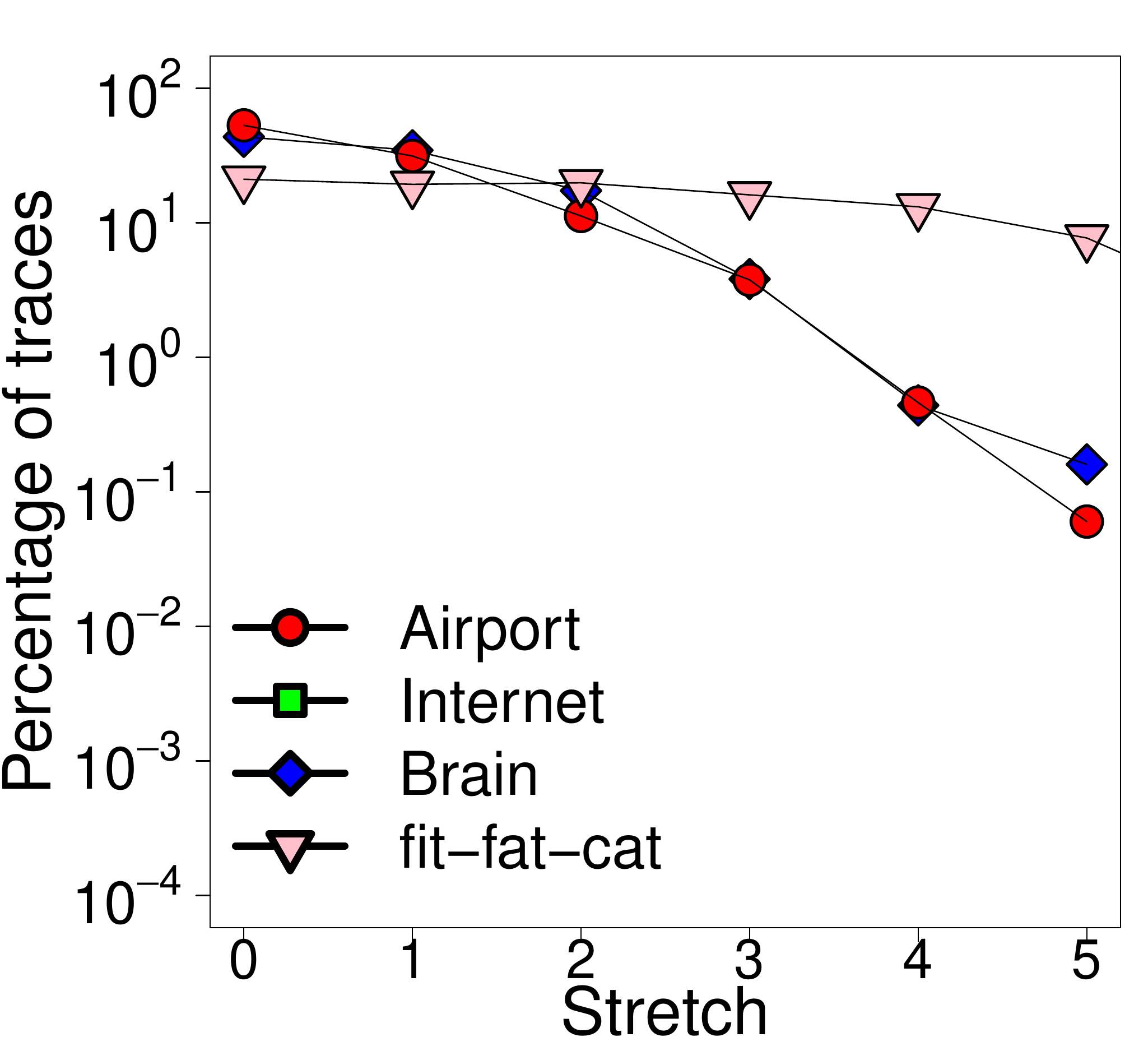}
      \put(-163.5,130.5){\large (\textbf{b})}
    \end{subfigure}
  }
  \centerline{
    \begin{subfigure}[b]{0.35\textwidth}
      \includegraphics[width=\textwidth]{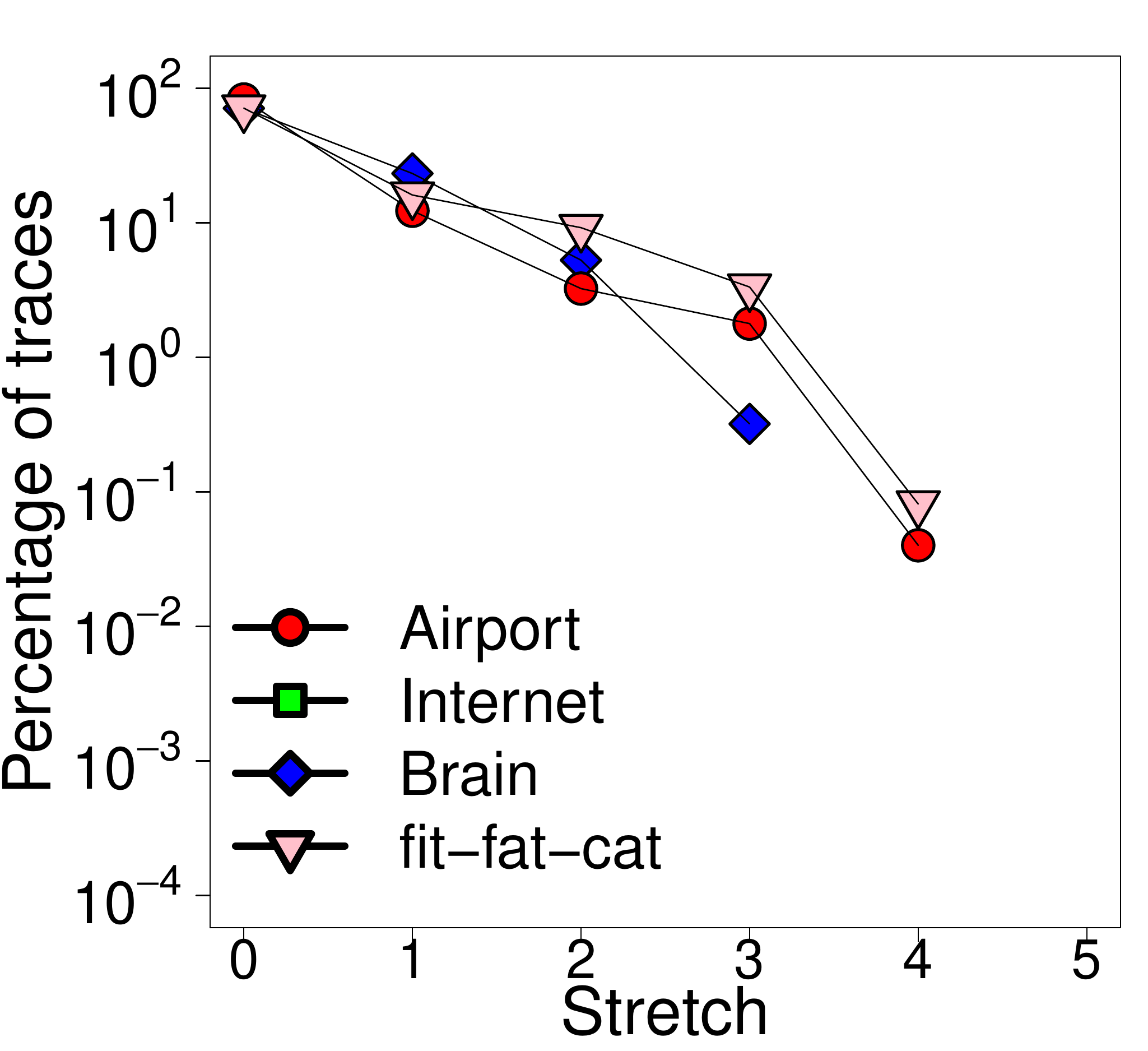}
      \put(-163.5,130.5){\large (\textbf{c})}
    \end{subfigure}
    \begin{subfigure}[b]{0.35\textwidth}
      \includegraphics[width=\textwidth]{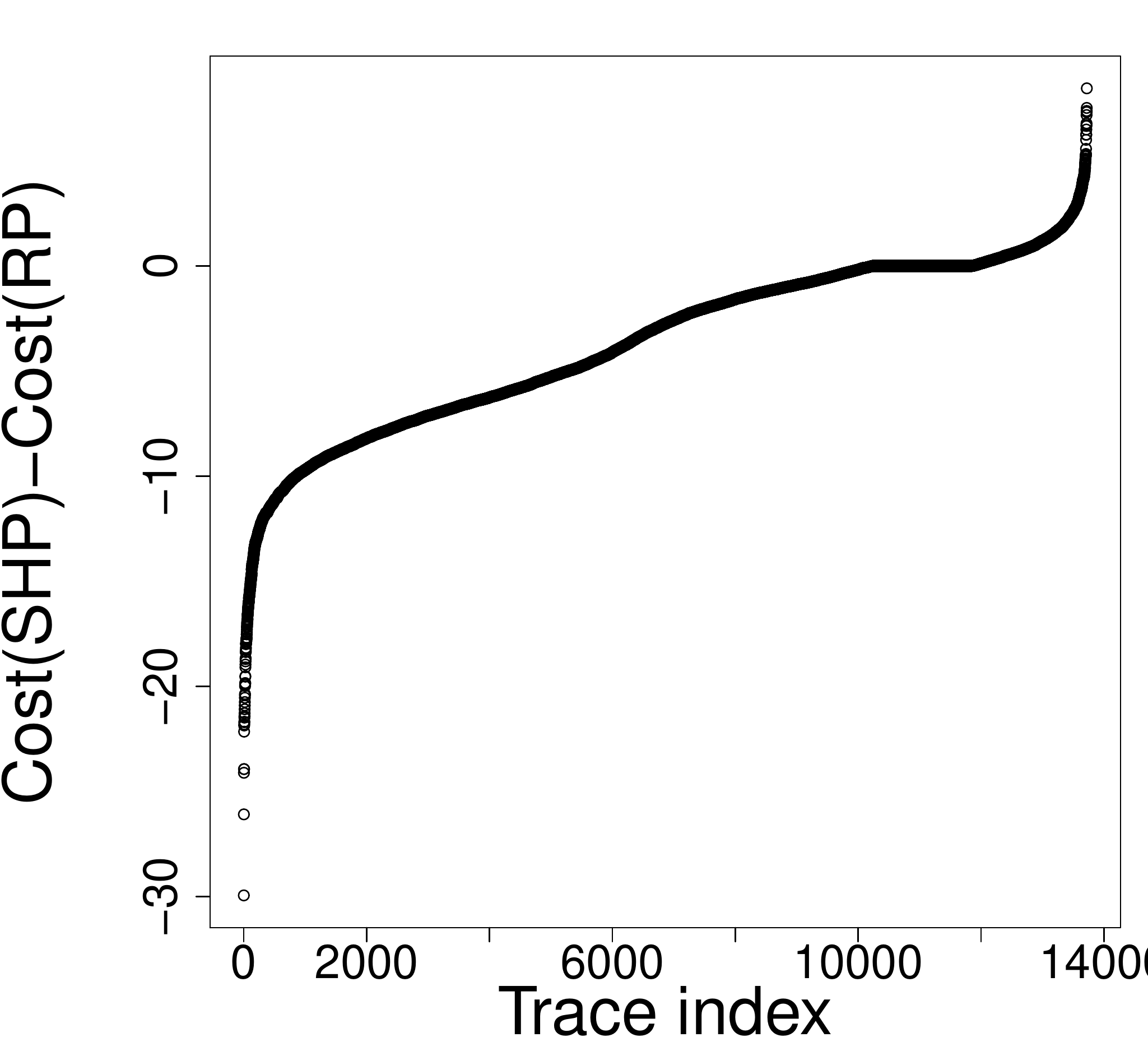}
      \put(-153.5,130.5){\large (\textbf{d})}
    \end{subfigure}
  }
  \caption{Results with betweenness based policies. Panel (\textbf{a})
    replicates the stretch data from real networks. Panel (\textbf{b})
    shows the stretch results of an experiment where all link weights
    are set to $\frac{1}{\text{EB}_i}$ and shortest paths
    are generated using these weights. Stretch distribution by using
    $\log(\frac{\max(\text{EB})}{\text{EB}_i})$
    as link weights is presented in panel (\textbf{c}). This plot
    matches better with the real data on panel (\textbf{a}). However,
    panel (\textbf{d}) indicates that this setting cannot be the cause
    of stretch in real networks, since in this weighting the sum of
    weights on the real paths is larger then of the simple hop-based
    shortest paths where all weights are set to one.}
   \label{fig:5}
\end{figure}

The researchers' desire towards simplicity would suggest that there
should be a simple weighting of the edges over which shortest path
computation can recover all the observed statistics of real
paths. Although we cannot disprove that such weighting exists, we list
here some of our experiments to illustrate that obtaining the correct
weights is far from trivial. In our first attempt we set the weights
to $w_i = \frac{1}{\text{EB}_i}$, where EB stands for egde
betweenness. This setting reflects the intuition that edges with high
betweenness are ``cheaper'' or ``less congested'' or ``have more
capacity'' and using them is better than using hop based shortest
paths (where $w_i = 1$). Panel (\textbf{b}) of Figure~\ref{fig:5}
shows that computing the shortest paths over this weighting results in
the appearance of stretch indeed. However the stretch in this case is
way higher than we experience in real networks.  In this setting
$50-60$\% of the paths are inflated in contrary with our real data
which exhibit stretch only for $\approx 30$\% of the real paths. Even
the decay of the stretch distribution is clearly different than the
real data replicated in panel (\textbf{a}) of Figure~\ref{fig:5}. We
have also tried the setting $w_i = \frac{1}{\sqrt{\text{EB}_i}}$ which
produces a very similar plot.  

We could recover very realistic stretch distribution by applying the
weighting: $w_i=\log(\frac{\max(\text{EB})}{\text{EB}_i})$ (see panel
(\textbf{c}) in Figure~\ref{fig:5}). This setting is based on the
observation that the availability of edges with high betweenness is
larger i.e. edges in the core are more reliable. Computing shortest
paths over these weights gives a path that is available with the
highest probability. Although the stretch distribution is very
promising in this case panel (\textbf{d}) points that such weighting
cannot be the reason for stretch in real networks. The plot shows
$\sum_{i \in SHP}w_i - \sum_{i \in RP}w_i$ for all source target pairs of
our real traces, where SHP stands for the hop based shortest path
using $w_i=1$ and RP stands for the real paths. In $80$\% of the cases
the real path is longer that the hop based shortest path (where the plot
is negative). This means that the hop based shortest path is actually
better in this weighting that the real paths.  

\end{document}